\newif\ifisonecolumn
\isonecolumntrue 
 
\ifisonecolumn
\documentclass[onecolumn,draft,12pt]{IEEEtran}
\else
\documentclass{IEEEtran}
\fi
\ifisonecolumn
\newcommand{\TwoOneColumnAlternate}[2]{#2}
\else
\newcommand{\TwoOneColumnAlternate}[2]{#1}
\fi
\usepackage{etoolbox}
 
\newcommand{\FigDat}[2]{
\ifstrequal{#1}{System}{
\begin{figure}[t]
        \includegraphics[scale=#2]{Fig_system}
    \caption{System model. .}
    \label{f:system model}
\end{figure}
}{}
\ifstrequal{#1}{Performance}{
\begin{figure}[t]
\end{figure}
}{}
}
\TwoOneColumnAlternate{
}{}

\usepackage[final]{graphicx}
\usepackage{amsmath,epsfig,amssymb}
\usepackage{nicefrac}
\usepackage{bm}
\usepackage{cite}

\begin{document}

\title{Variable pool testing for infection spread estimation}

\author{Itsik Bergel}

\maketitle
\begin{abstract}
We present a method for efficient estimation of the prevalence of infection in a population with high accuracy using only a small number of tests.  The presented approach uses pool testing with a mix of pool sizes of various sizes. The test results are then combined to generate an accurate estimation over a wide range of infection probabilities. This method does not require an initial  guess on the infection probability. We show that, using the suggested method, even a set of only $50$ tests with a total of only $1000$ samples can produce reasonable estimation over a wide range of probabilities. A measurement set with only $100$ tests is shown to achieve $25\%$ accuracy over infection probabilities from $0.001$ to $0.5$. The presented method is applicable to COVID-19 testing.

\end{abstract}
\section{Background}\label{sec:sysmodel}
In times of epidemic crisis, it is important to monitor the spread of the infection in the general population as well as in various sub-groups (neighborhoods, work places, special populations etc.). While taking samples can be relatively easy, the actual tests require sophisticated machinery and specific reagents which are often sparse. Furthermore, the testing procedure is time consuming, and the processing of each test may take several hours. Thus, most countries focus their testing capacity on  acutely ill patients, leaving very small capacity to statistical studies of other populations. Furthermore, the prevalence disease carriers in such other population can be low, and hence, a direct estimation would require many tests.

Pool testing (or group testing) was suggested as a method to efficiently test large populations \cite{dorfman1943detection,bilder2012pooled}.  In pool testing several samples are mixed
and tested at a single pool. That is, A group of collected samples (e.g., using swabs for COVID-19 tests) is mixed into a single tube. This tube is then tested (e.g., using RT-qPCR). Thus, while many samples are collected, the load on the testing machinery remains small.

Pool testing is used for various infectious disease \cite{litvak1994screening,nguyen2019methodology}, and was proven to work also for RT-qPCR tests \cite{taylor2010high,arnold2013evaluation}. Furthermore, pool testing was successfully demonstrated recently for the SARS-CoV-2  pathogen of  COVID-19 \cite{yelin2020evaluation} for pool sizes of up to $64$ samples. Pool sampling strategics for COVID-19 (with fixed pool size) where also studied in \cite{sinnott2020evaluation}.

The literature on pool testing is quite diverse, both in the medical literature (e.g., \cite{schmidt2010blood,wang2015general}) and in data processing contexts (e.g., \cite{atia2012boolean,bai2019adaptive}).  Pool testing is commonly used for efficient detection of infected samples when the infection probability is small. But, pool testing is also used for efficient estimation of
the prevalence of a rare disease. 

But, using pool testing for prevalence estimation, requires a choice of the pool size in accordance with the expected prevalence. This is problematic, as such tests are often performed without prior knowledge on the tested group. For example  \cite{nguyen2018sequential} suggested a sequential pool testing, where at each stage the estimate of the infection probability is improved and used to design the next stage. Yet, the test accuracy still strongly depends on the quality of the initial guess.

So far, there is no simple and efficient method to choose pool sizes that will bring accurate estimation over a wide range of infection probabilities in a single batch.

In this work we present a method for efficient estimation of the prevalence of infection in a population using a small number of tests. The presented approach uses a mix of pool sizes, ranging from single sample test to very large pools. The test results are then combined to generate an accurate estimation for a wide range of infection probabilities. This solves the problem of choosing the pool size (which requires an initial guess of the probability). As an example, using only $100$ tests, we can estimate the infection probability at an accuracy of $\pm25$\% over all the probability range from $10^{-3}$ to $0.5$.    

In the following we first consider pool testing with fixed pool size, and then present the variable pool size approach. 
\section{performance analysis for constant pool size}

We consider the estimation of the spread of disease in a given population. Denote the population size by $L$ and the number of infected  by $L_\mathrm{i}$. We define the probability of finding an infected sample by: 
$p=\frac{L_\mathrm{i}}{L}
$. We consider the estimation accuracy given a limited number of tests $T$. 

For the initial step, we assume that all pools are of equal size $N$. A pool test is positive if at least one of its samples is positive. Thus, $\delta_i$, the result of test $i$ will have a binary distribution, with:
\begin{IEEEeqnarray}{rCl}\label{XXX}
\Pr (\delta_i=0)=(1-\Pr (\delta_i=1))=(1-p)^N.
\end{IEEEeqnarray}
Note that $\delta_i=1$ indicates a positive test for the disease. 

Considering a maximum likelihood (ML) estimator, we have:
\begin{IEEEeqnarray}{rCl}\label{e:ML_hom}
\hat p &=& \arg\max_p \prod_{i=1}^T (1-p)^{N(1-\delta_i)}(1-(1-p)^N)^{\delta_i}
\end{IEEEeqnarray}
Let $w=\sum_{i=1}^T \delta^i$, we take the derivative of the log of \eqref{e:ML_hom} with respect to $p$, and compare to zero:
\begin{IEEEeqnarray}{rCl}\label{XXX}
0&=&-N(T-w)\frac{1}{1-\hat{p}}+w\frac{N(1-\hat{p})^{N-1}}{1-(1-\hat{p})^N}
\end{IEEEeqnarray}
which is solved by:
\begin{IEEEeqnarray}{rCl}\label{XXX}
\hat{p}=1-\left(1-\frac{w}{T}\right)^{1/N}
\end{IEEEeqnarray}

\begin{figure}
    \center
\includegraphics{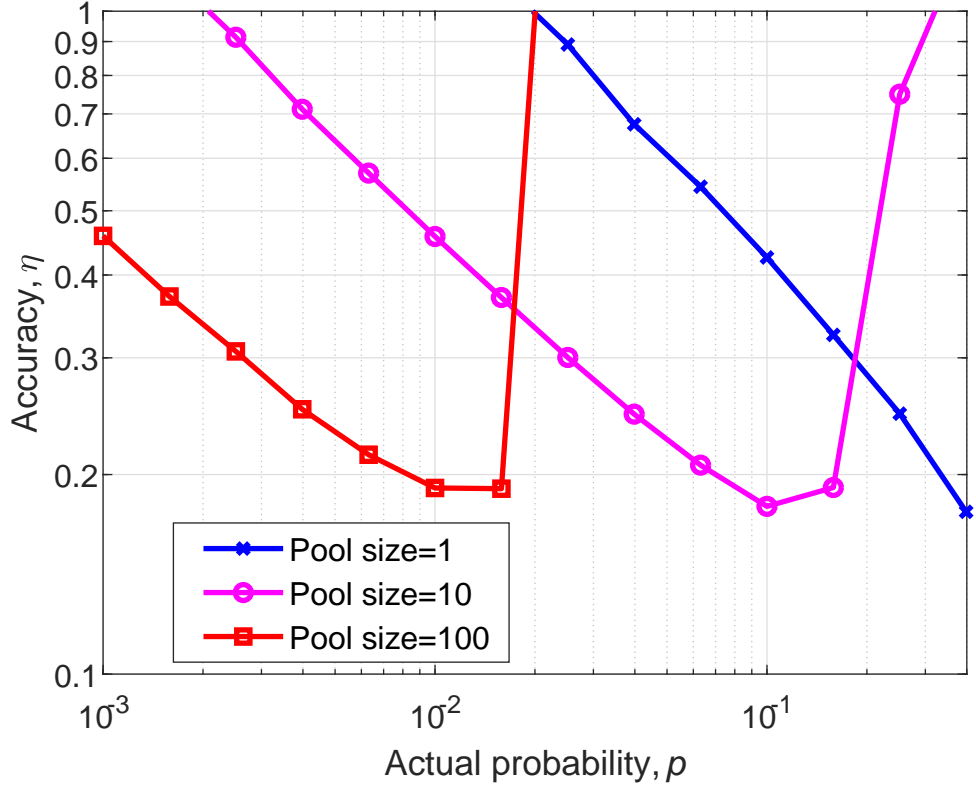}
\caption{Estimation accuracy vs. the actual infection probability for various pool sizes. The estimation accuracy is defined in \eqref{eq:estimation_accuracy}. (Recall the lower values of $\eta$ represent better accuracy.)}
\label{fig:acc_vs_N}
\end{figure}

The accuracy of this estimation is presented in figure \ref{fig:acc_vs_N}.
We measure the accuracy relative to the actual infection probability. The root mean square error is defined as:
\begin{IEEEeqnarray}{rCl}\label{XXX}
\mbox{RMSE}&=&\sqrt{E\left[\left(\hat{p}-p\right)^2\right]}
\end{IEEEeqnarray}
and is evaluated using Monte Carlo simulation.
The estimation accuracy is defined as 
\begin{IEEEeqnarray}{rCl}\label{eq:estimation_accuracy}
\eta=\frac{\mbox{RMSE}}{p}.
\end{IEEEeqnarray}
Note that the accuracy is better if $\eta$ is smaller. All Monte Carlo simulations in this work use $10^4$ repetitions. 

Fig. \ref{fig:acc_vs_N} show the accuracy as a function of the actual infection probability for pool sizes of $10$ and $100$ as well as without pooling (pool size of 1). The figure demonstrates that pooling can significantly improve the accuracy. But, it is useful  only if the pool size is approximately $1/p$. If the pool size is too small, then the probability to get a positive set is still too small and the efficiency is reduced. On the other hand, if the pool size is too large, almost all tests will turn positive and the accuracy degrades very fast.  

Thus, to have an efficient test, one must match the pool size to an initial guess of the actual probability. To avoid the need for such a guess, the next sub-section presents an efficient method for estimation with variable pool sizes.

\section{General pool sizes}
\subsection{Probability estimation}
We next consider a general pooling scheme where the pool of test $i$ is of size $N_i$.  Thus, $\delta_i$, is a binary distribution, with:
\begin{IEEEeqnarray}{rCl}\label{XXX}
\Pr (\delta_i=0)=(1-\Pr (\delta_i=1))=(1-p)^{N_i}
\end{IEEEeqnarray}
(and again,  $\delta_i=1$ indicates a positive test for the disease). 

The ML estimate of the infection probability, $p$, from a set of $T$ tests with pool sizes $N_1,N_2,\ldots,         N_T$ is given by:
\begin{IEEEeqnarray}{rCl}\label{e:ML_gen}
\hat p &=& \arg\max_p \prod_{i=1}^T (1-p)^{N_i(1-\delta_i)}(1-(1-p)^{N_i})^{\delta_i}
\end{IEEEeqnarray}
Taking the derivative of the log of \eqref{e:ML_gen}, and comparing to zero:
\begin{IEEEeqnarray}{rCl}\label{XXX}
0&=&\sum_{i=1}^TN_i\frac{1}{1-\hat{p}}(1-\delta_i)-\delta_i\frac{N_i(1-\hat{p})^{N_i-1}}{1-(1-\hat{p})^{N_i}}
\\&=&\sum_{i=1}^TN_i(1-\delta_i)-\delta_i\frac{N_i(1-\hat{p})^{N_i}}{1-(1-\hat{p})^{N_i}}.
\end{IEEEeqnarray}
Thus, the ML estimator is the solution to 
\begin{IEEEeqnarray}{rCl}\label{eq:Gen_pool_MLE}
\sum_{i=1}^T N_i\left(1-\frac{\delta_i}{1-(1-\hat{p})^{N_i}}\right)=0 .
\end{IEEEeqnarray}
In this case, the ML estimator does not have a closed form expression. Yet, the left hand side of Equation \eqref{eq:Gen_pool_MLE} is monotonic increasing with $\hat{p}$. Hence,   the ML estimator can be efficiently calculated by solving \eqref{eq:Gen_pool_MLE} using a binary search.

\subsection{Choosing the pool sizes}
As shown above, the probability estimate will benefit most from a pool size which is approximately $N=1/p$. In this approach we wish to measure a large range of possible infection probabilities, $p$. Thus, we need to use pool size with wide range of sizes. To do so, we suggest to select the pool sizes in a logarithmic manner, that is:
\begin{IEEEeqnarray}{rCl}\label{XXX}
N_i=\lceil N_0 \cdot q^i\rfloor
\end{IEEEeqnarray}
for $i=0,\ldots, T-1$, where $N_0$ is the size of the smallest pool and $q>1$ is the logarithmic spacing. We use the notation $\lceil \cdot\rfloor$ to indicate rounding to the nearest integer.

\begin{figure}
\center
\includegraphics{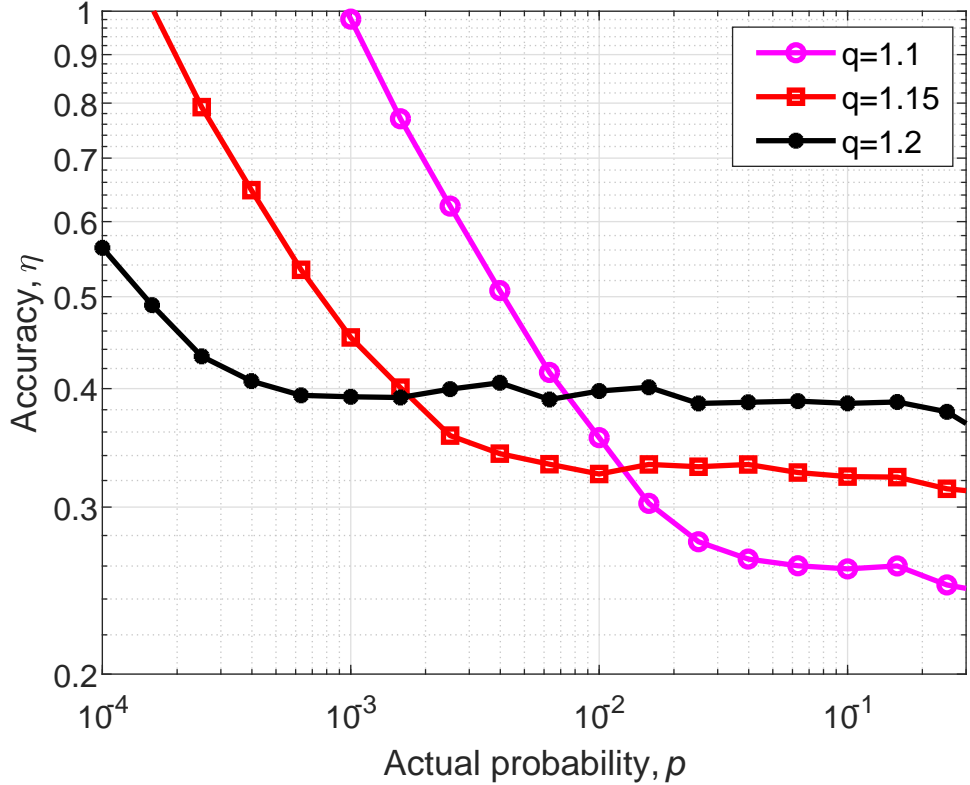}
\caption{Estimation accuracy vs. the actual infection probability for various $q$ values.}
\label{fig:acc_vs_q}
\end{figure}

If the desired range of measure probabilities is $p_{\min}<p<p_{\max}$, then it is important to have $N_0<1/p_{\max}$ and $N_0\cdot q^{T-1}>p_{\min}$. Thus, the choice of $q$ represents a tradeoff between a large measurement range and the measurement accuracy. This is demonstrated in Fig. \ref{fig:acc_vs_q}, where the estimation accuracy, $\eta$, is depcited as a function of the actual infection probability for $T=50$ tests, $N_0=1$ and $q=1.1$, $q=1.15$ and $q=1.2$. The figure shows the the logarithmic choice of pool sizes indeed keeps the estimation accuracy nearly constant, but only within the dynamic range of the measurements. For the values of this figure we have $N_0\cdot q^{T-1}=107, 942$ and $7584$ for the values of $q$ give above. Indeed, we see that the estimation accuracy starts to deteriorate around $1/(N_0\cdot q^{T-1})$, that is around $0.009$, $0.001$ and $0.00013$ respectively.
Note that even using the larger $q$ in this simulation, the accuracy of $\pm40\%$ is quite good, as this error is obtained in a measurement that covers three order of magnitudes of the actual probability.

\subsection{Practical example}
\begin{figure}
\center
\includegraphics{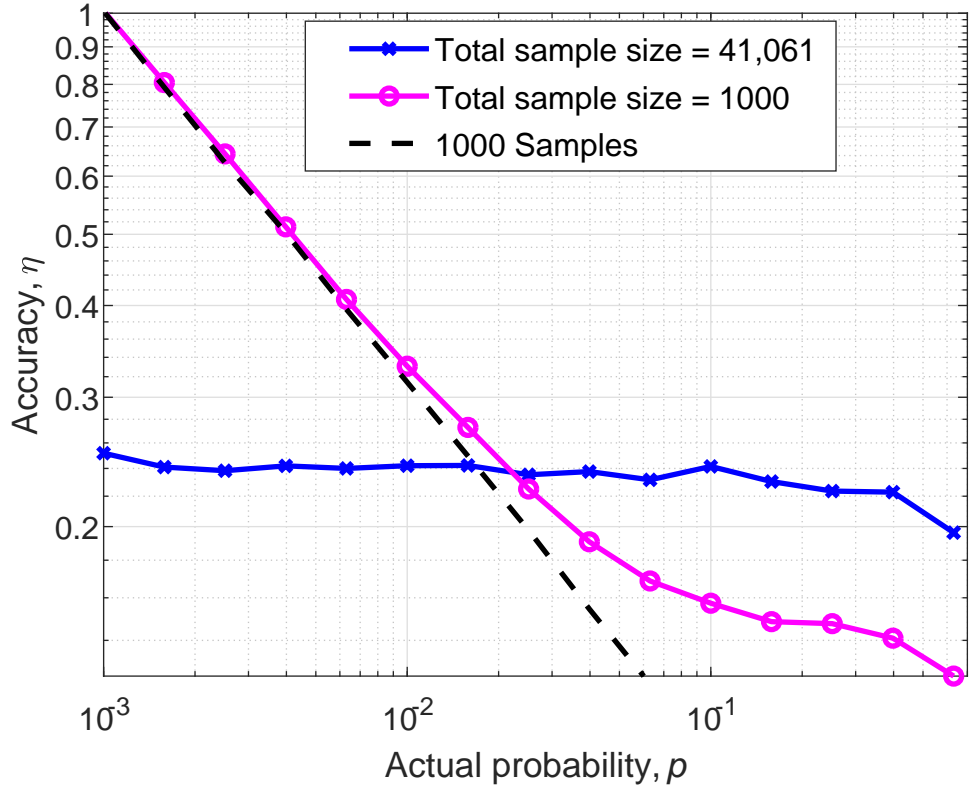}
\caption{Estimation accuracy vs. the actual infection probability. The solid lines uses $100$ samples, one is optimized for accuracy, while the other is limited to $1000$ patients. The dashed line uses $1000$ samples and performs individual test for each sample.}
\label{fig:acc_vs_sample}
\end{figure} 

We next present a last set of simulations that demonstrates the efficiency of the suggest approach for a fast and efficient measurement of the infection probability. We choose the measurement design to cover the range $10^{-3}<p<0.5$. Using $T=100$ tests, we use $N_0=1$ and $q=1.085$ (such that $q^{T-1}=10^{3.5})$. The resulting accuracy is depicted by the solid line with $x$-markers in Fig. \ref{fig:acc_vs_sample}. The figure demonstrates that using only $100$ tests, we can get an estimation accuracy of $25\%$ over the whole probability range.

The main drawback of this approach is that it requires many samples. The example above used $100$ tests, but required a total of $40,439$ samples (mixed in the various pools). This is sometimes a problem, as it may be difficult to persuade these many people to come to test. Thus, the figure also depicts the accuracy when the measurement is limited only to $1000$ samples.  In this case, we adjust the $q$ to achieve a logarithmic spacing such that the sum of all pool sizes in $1000$ (i.e., $q=1.03708$). 

Such a limiting of the total number of tests obviously limits the capability of measuring very low infection probabilities. The resulting accuracy is shown by the solid line with o-markers. The use of smaller pool sizes improves the accuracy at higher infection probabilities, but is indeed limited at low infection probabilities. As a comparison, we plot also the estimation accuracy of the conventional approach, i.e., taking $1000$ samples and performing individual test to each sample ($1000$ tests). This conventional approach requires much more resources. Yet, it gives significant advantage  only if the infection probability is above $0.05$.

\begin{figure}
\center
\includegraphics{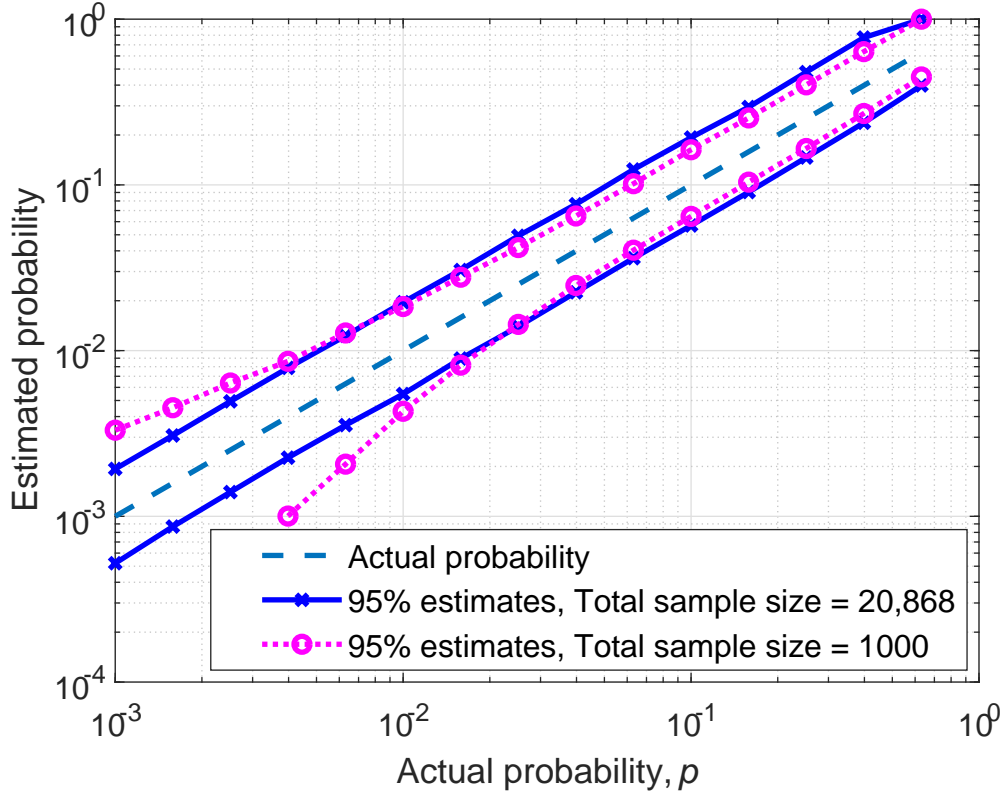}
\caption{$95\%$ confidence interval for practical estimation with $T=50$ tests.}
\label{fig:Conf95_vs_p}
\end{figure} 

Fig. \ref{fig:Conf95_vs_p} depicts the performance for a similar scenario, but using only $50$ tests. The figure shows the performance using our suggested pool sizes ($N_0=1$, $q=1.1788$ total of $20,868$ samples), and the performance with using only $1000$ samples ($N_0=1$, $q=1.09578$).
The figure show the $95\%$ confidence interval of the estimation for each actual infection probability. The figure shows that even only $50$ tests can give reasonable estimations when the pool size is not limited. Even with a limit of only $1000$ samples, the estimation deteriorates only below $p=10^{-2}$. Moreover, even in this case, the estimates still give a good upper-bound on the infection probability.  

\section{Conclusions and future work}
We presented a method for efficient estimation of infection probability using a small number of tests. The method is based on pool testing with variable pool sizes. It is shown that proper choice of pool sizes leads to accurate estimations even with small number of tests. For example, using $100$ tests was shown to achieve $25\%$ accuracy over a wide range of actual infection probabilities. Even a set of only $50$ tests over only $1000$ samples was shown to produce reasonable estimation.

Further research is required in order to accommodate for false alara (false positive) and miss detection (false negative) probabilities. In particular, it can be assumed that these error probability can increase with the pool size and hence effect the choice of pool sizes. Pool tests for COVID-19 was so far demonstrated for up to $64$ samples in a pool, with small enough errors \cite{yelin2020evaluation}. The data in that work can give an initial estimate on the behavior of the error probabilities, and hence used to improve the estimation and the pool size design.

\newcommand\comment[1]{}
\comment{
\begin{appendices}
\section{Analytic optimization for constant pool size}\label{app:pool_size}
Testing the estimate accuracy, we aim for an error of $\pm d$\%. That is, we wish to evaluate the probability:
\begin{IEEEeqnarray}{rCl}\label{XXX}
p_\mathrm{acc}=\Pr(b_\mathrm{L}\cdot p<\hat p<b_\mathrm{H}\cdot p)
\end{IEEEeqnarray}
where $b_\mathrm{L}=1-d/100$ and $b_\mathrm{L}=1+d/100$.  
We have 
\begin{IEEEeqnarray}{rCl}\label{XXX}
p_\mathrm{acc}=\Pr\left(T\left(1-(1-b_\mathrm{L}\cdot p)^N\right)< w<T\left(1-(1-b_\mathrm{H}\cdot p)^N\right)\right)
\end{IEEEeqnarray}
where $w$ is a binomial distribution
with parameters $T$ and $1-(1-p)^N$. If $w\gg 1$, is is reasonable to approximate the distribution of  $w$ as a Gaussian distribution with mean $T-T(1-p)^N$ and variance $T(1-(1-p)^N)(1-p)^N$. Thus, we can approximate:
\begin{IEEEeqnarray}{rCl}\label{XXX}
\Pr\left( w>T\left(1-(1-b_\mathrm{H}\cdot p)^N\right)\right)&=&Q\left(\frac{T\left(1-(1-b_\mathrm{H}\cdot p)^N\right)-T+T(1-p)^N}{\sqrt{T(1-(1-p)^N)(1-p)^N}}\right)
\\&=&Q\left(\frac{-(1-b_\mathrm{H}\cdot p)^N+(1-p)^N}{\sqrt{(1-(1-p)^N)(1-p)^N}}\sqrt{T}\right)
\end{IEEEeqnarray}
To maximize the accuracy, we take the derivative of the internal part with respect to $N$ and compare to zero:
\begin{IEEEeqnarray}{rCl}\label{XXX}
0&=&\frac{(-(1-b_\mathrm{H}\cdot p)^N\log(1-b_\mathrm{H}\cdot p)+(1-p)^N\log(1-p))\sqrt{(1-(1-p)^N)(1-p)^N}}{(1-(1-p)^N)(1-p)^N}
\\&&-\frac{(-(1-b_\mathrm{H}\cdot p)^N+(1-p)^N)\frac{(1-(1-p)^N)(1-p)^N}{2\sqrt{(1-(1-p)^N)(1-p)^N}}}{(1-(1-p)^N)(1-p)^N}
\end{IEEEeqnarray}

 we have:
\begin{IEEEeqnarray}{rCl}\label{XXX}
E[\hat{p}]=\frac{E[w^{1/N}]}{T^{1/N}}
\end{IEEEeqnarray}
\begin{IEEEeqnarray}{rCl}\label{XXX}
RMSE&=&\sqrt{E\left[\left(\hat{p}-p\right)^2\right]}=\sqrt{E\left[\left(\frac{w^{1/N}}{T^{1/N}}-p\right)^2\right]}
\\&=&\sqrt{E\left[\left(\frac{(T-\bar w)^{1/N}}{T^{1/N}}-p\right)^2\right]}
\\&=&\sqrt{E\left[\left((1-\frac{\bar w}{T})^{1/N}-p\right)^2\right]}
\end{IEEEeqnarray}
if $\bar w\ll T$ then:
\begin{IEEEeqnarray}{rCl}\label{XXX}
RMSE&\approx&\sqrt{E\left[\left(1-\frac{\bar w}{N T}-p\right)^2\right]}
\\&=&\sqrt{E\left[(1-p)^2-2(1-p)\frac{\bar w}{N T}+(\frac{\bar w}{N T})^2\right]}
\\&=&\sqrt{E\left[(1-p)^2-2(1-p)\frac{\bar w}{N T}+(\frac{\bar w}{N T})^2\right]}
\end{IEEEeqnarray}
We use $E[\bar w]=T-E[w]=T(1-p^N)$, and $\mbox{Var}(\bar w)=\mbox{Var}(w)=T(p^N-p^{2N})$, $E[\bar w^2]=E^2[w]+\mbox{Var}(w)=T^2(1-p^N)^2+T(p^N-p^{2N})$
\begin{IEEEeqnarray}{rCl}\label{XXX}
RMSE&\approx&\sqrt{(1-p)^2-2(1-p)\frac{T(1-p^N)}{N T}+\frac{T^2(1-p^N)^2+T(p^N-p^{2N})}{N^2 T^2}}
\\&=&\sqrt{(1-p)^2-2(1-p)\frac{(1-p^N)}{N }+\frac{(1-p^N)^2}{N^2 }+\frac{(p^N-p^{2N})}{N^2 T}}
\\&=&\sqrt{(1-p-(1-p^N))^2+\frac{(p^N-p^{2N})}{N^2 T}}
\\&=&\sqrt{(p^N-p)^2+\frac{(p^N-p^{2N})}{N^2 T}}
\end{IEEEeqnarray}

\end{appendices}
}
\bibliographystyle{IEEEtran}
\bibliography{bib_cor}
\end{document}